\PassOptionsToPackage{table,xcdraw}{xcolor}
\documentclass[manuscript,screen,nonacm]{acmart}

\usepackage{hyperref}
\usepackage{hyperxmp}

\copyrightyear{2025}
\acmYear{2025}

\begin{document}

\title[Exploring the Use of Predictive Analytics by Austrian Tax Authorities]{Exploring the Use of Predictive Analytics by Austrian Tax Authorities: \\ A Qualitative Study within the Task--Technology Fit Model}

\author{Simon Staudinger}
\orcid{0000-0002-8045-2239}
\email{simon.staudinger@jku.at}
\affiliation{%
  \institution{Institute of Business Informatics -- Data \& Knowledge Engineering, Johannes Kepler University Linz}
  \streetaddress{Altenberger Str. 69}
  \city{Linz}
  \country{Austria}
  \postcode{4040}
}

\author{Christoph G. Schuetz}
\orcid{0000-0002-0955-8647}
\email{christoph.schuetz@jku.at}
\affiliation{%
  \institution{Institute of Business Informatics -- Data \& Knowledge Engineering, Johannes Kepler University Linz}
  \streetaddress{Altenberger Str. 69}
  \city{Linz}
  \country{Austria}
  \postcode{4040}
}

\author{Marina Luketina}
\email{marina.luketina@jku.at}
\orcid{0009-0007-7440-6002}
\affiliation{%
  \institution{University of Applied Sciences Upper Austria}
  \streetaddress{Wehrgrabengasse 1-3}
  \city{Steyr}
  \country{Austria}
  \postcode{4400}
}

\begin{abstract}
Taxes finance important government services that are now taken for granted in our society, such as infrastructure, health care, or retirement pensions. 
Tax authorities everywhere strive to ensure that all individuals and organizations comply with applicable tax laws.
In this regard, tax authorities must prevent individuals and organizations from evading taxes in an illegal manner.
To this end, Austrian tax authorities employ state-of-the-art predictive analytics technology for the selection of suspicious cases for tax audits, thus making efficient use of scarce resources for tax auditing. 
In this paper, we explore how Austrian tax authorities employ predictive analytics technology in tax auditing and how well the use of such technology fits the characteristics of the task at hand.
We collaborated with the Austrian Federal Ministry of Finance's Predictive Analytics Competence Center to obtain insights into the application of predictive analytics technology by Austrian tax authorities. 
The thus obtained insights serve as the basis for a qualitative analysis in the context of the task-technology fit framework.

\end{abstract}

\ccsdesc[500]{Information systems~Data mining}
\ccsdesc[500]{Applied computing~Law}
\ccsdesc[500]{Information systems~Decision support systems}

\keywords{tax auditing, information systems, machine learning}

\maketitle

\section{Introduction}
Tax fraud and tax evasion pose significant challenges for governments striving to fund vital services for society, such as infrastructure development and upkeep, education, healthcare, and retirement benefits.
As an illustration, in 2017, the value-added tax (VAT) gap, which represents the difference between anticipated and realized VAT revenues, reached approximately €2.4 billion in Austria and a staggering €137 billion across the entire European Union~\cite{poniatowski2019study}.
While this gap is not only caused by tax fraud and tax evasion, but also includes other unforeseen events such as bankruptcies, tax fraud and tax evasion nevertheless constitute a considerable portion of the VAT gap.
The government is committed to upholding individuals' and organizations' compliance with tax laws to the greatest extent possible and to monitor adherence using the resources available to authorities.

Since 2014, the Austrian Federal Ministry of Finance (BMF) has been employing predictive analytics in order to identify suspected cases of tax fraud and illegal tax evasion, culminating in the foundation of the \emph{Predictive Analytics Competence Center}~(PACC) in 2016/17.
The BMF refers to predictive analytics as ``an approach used in data analysis'', which ``uses historical data to predict future events'', with the historical data being ``used to create mathematical models, which are then applied to current data''~\cite{pacc}.
This definition aligns with common definitions of the term ``predictive analytics'' found in research and practice. 
Commonly, different types of analytics are distinguished, among them often \emph{descriptive}, \emph{predictive}, and \emph{prescriptive} analytics.
While descriptive analytics comprises ``statistical methods designed to explore what happened'', predictive analytics comprises ``machine learning methods designed to predict what will happen next'', with prescriptive analytics referring to ``methods designed to answer what should the business do next''~\cite{mortenson2015}.
Similarly, Holsapple et al.~\cite{holsapple2014} define predictive analytics as ``predicting what may or will occur''.

In this paper, we investigate the use of predictive analytics by Austrian tax authorities for the purposes of detecting suspicious tax reports as the basis for selecting cases for tax auditing.
In particular, we identify task characteristics and describe the technology employed by the BMF's PACC, and we evaluate the fit between the task at hand and the employed technology.
The addressed research questions are, thus, as follows.

\begin{enumerate}
    \item \textit{How is predictive analytics used by Austrian tax authorities~(PACC) for the selection of cases for tax auditing?}

    \item \textit{Which factors determine the degree of fit between the task of selecting cases for tax auditing and predictive analytics technology?}
\end{enumerate} 
To answer the first question, we conducted interviews with experts of the BMF's PACC to explore how predictive analytics is used by Austrian tax authorities. 
To answer the second question, we apply the task--technology fit framework~\cite{goodhue1995task} to the insights obtained from the interviews.
Furthermore, we conducted a survey among tax auditors, which together with the interviews of the PACC experts served as the basis for a qualitative content analysis.

The remainder of this paper is structured as follows.
In Section~\ref{sec:related} we review related work.
In Section~\ref{sec:methodology} we briefly describe the employed methodology.
In Section~\ref{sec:aspects} we describe the task, technology, and legal aspects.
In Section~\ref{sec:fit}, we investigate the task--technology fit.
In Section~\ref{sec:conclusion}, we conclude this paper with the implications for research and practice as well as an outlook on future work.

\section{Related Work}\label{sec:related}

Past work has highlighted the potential that the use of data holds in the public sector in all sorts of domains.
Gamage \cite{gamage2016new} highlights that the ever-growing amount of data changes the way how people work and live.
Private sectors like logistics, retail, and financial services are already leveraging data analytics to achieve benefits compared to their competitors who rely on solely experience and expertise. 
It was proposed that these benefits might also be achieved for applications in the public sector. 
Gamage~\cite{gamage2016new} describes cases of the public sector successfully leveraging big data and data analytics for various tasks, highlighting also the challenges faced by the public sector.

Kim et al.~\cite{kim2014big} analyze similarities and dissimilarities between the business sector and the public sector regarding the applications of big data. 
Despite the differences in the primary missions of businesses and public sector---businesses aim for generating profit whereas the public sector strives for sustainable development, securing citizens' fundamental rights, and promoting general welfare---there is no doubt in the potential added value through the use of predictive analytics.
Kim et al. mention that collecting and using big data for predictive analytics and ensuring basic rights of privacy might be challenging.

Previous works in the tax domain highlighted the importance and possible value that can be generated by the use of predictive analytics.
Mehta et al.~\cite{DBLP:conf/bigdataconf/MehtaBRK20} trained a neural network to predict potential tax return defaulters in India with the goal to achieve cost savings.  
Data that was used to train the neural network was provided by the government of Telangana, India, and includes information about the social environment of the taxpayer, the number of late filed tax returns, the geographic division of the taxpayer, the average paid tax per month, and the total amount of purchases.
The authors proposed the use of a cost function which also takes into account the cost of possible misclassification instead of simply minimizing the total number of misclassifications. 

Göschlberger \& Deliu \citep{DBLP:conf/snams/BernhardD21} assessed the use of different approaches to apply social network analysis algorithms for fraud prediction at the Austrian Federal Ministry of Finance. 
The authors compared results regarding the ranking of companies based on known risk-information given by the PageRank algorithm and a personalized BiRank algorithm. 
The BiRank algorithm was shown to provide more accurate results on the given use case. 
Göschlberger~\& Deliu state that further work to employ more advanced versions of the network analysis might also yield fruitful results regarding fraud prediction.

\section{Methodology}\label{sec:methodology}
\begin{figure}[h]
  \centering
  \includegraphics[width=\linewidth]{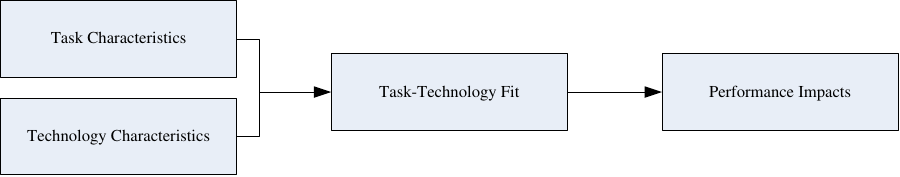}
  \caption{Task--technology fit model~\citep{goodhue1995task}}
  \label{figure_ttf}
\end{figure}

We conduct an exploratory study using the task--technology fit (TTF) framework~\cite{goodhue1995task} as a conceptual basis to investigate the use of predictive analytics by Austrian tax authorities for the selection of cases for tax auditing.
The TTF framework is a widely used and proven framework to assess the link between information systems and the individual performance of its users. 
Figure~\ref{figure_ttf} illustrates the basic TTF model.
Goodhue~\& Thompson~\cite{goodhue1995task} proposed the TTF framework, asserting that whether a used technology has a positive impact on the performance can be measured by analyzing the fit of the technology with the tasks it should support.
The TTF framework assumes that there is no system that initially provides perfect data to meet complex task needs.
As the overall fit increases, the discrepancies between task requirements and the system's functionalities diminish. 
Performance impact is related to the accomplishment of a portfolio of tasks performed by an individual. 
To analyze the fit between the technology and the tasks, the characteristics of both need to be examined and further discussed why and how the technology can provide the necessary fit to support the execution of the tasks. 
Besides the fit, Goodhue~\& Thompson propose a second factor, \emph{utilization}, to impact the performance of information systems.

An extensive body of research has studied the TTF model~\cite{DBLP:journals/jcis/CaneM09,9535028}, including both quantitative and qualitive studies. 
Previous works~\citep{DBLP:journals/ijecommerce/GebauerS04,info:doi/10.2196/18414,DBLP:conf/ecis/SturmP20} employed the TTF model in exploratory case studies to assess the impact of the employed technology on the performance of the analyzed tasks.
In a similar vein, in this paper, we investigate the impact of predictive analytics technology on the task of tax auditing in the case of the BMF's PACC.

We collected data in two ways.
First, we conducted two interviews with experts from the BMF's PACC to gain a better understanding of working practices, task, and technology in the context of the identification of suspicious tax reports.
The information from the interviews provides the required basis for presenting the technological aspects of fraud detection in Section~\ref{sec:techAspects}, which is further used within the TTF framework to identify technology characteristics and fit factors.
We recorded and transcribed the interviews, excluding greetings and closure, for the purposes of analyzing the current state of operations at PACC.
For the transcripts of the interviews, due to the sensitive nature of the subject, we anonymized any names and other personal identifiers to protect individuals and companies. 
Furthermore, we conducted a survey among tax auditors who are responsible for conducting the actual tax audits, asking 13 questions related to the use of predictive analytics technology to support the selection of audit cases.
The survey was sent to 146 auditors, based in offices in Vienna or Graz. 
We received 31 answers, corresponding to a response rate of 21.2\%, which we evaluated qualitatively to incorporate the answers into our analysis of the PACC's activity in the context of the TTF framework. 
Specifically, we conducted qualitative content analysis across all survey questions according to Mayring~\cite{mayring2014} to identify task characteristics, technology characteristics, and the effect of those characteristics on the performance of the tasks.

\section{Task, Technology, and Legal Aspects}
\label{sec:aspects}
The Predictive Analytics Competence Center (PACC) was founded in the years 2016/17 and its main purpose is the support of governmental institutions by providing state-of-the-art prediction and analysis tools.
PACC's application areas are manifold, ranging from risk-based audit case selection to real-time risk assessment in multiple areas like tax or customs. 
In the following, we describe the task as well as the technological and legal aspects of the application of predictive analytics in tax auditing, focusing particularly on case selection for company audits and missing trader fraud detection.

\subsection{Task Description}
Tax evasion is a major challenge for tax authorities in countries all over the world. 
The loss from tax irregularities was estimated at \$483 billion worldwide in 2021 alone~\citep{globalalliance}. 
Tax evasion reduces the tax base and related public resources for the provisioning of public goods and erodes fiscal equity. 
In Austria, tax authorities have a legal mandate to combat tax evasion and avoidance (for further details see Section~\ref{sec:legal_aspects}).
One method applied by the Austrian Ministry of Finance to address this issue and prescribed by Austrian Procedural Tax Law (Section~147) is the tax audit---an examination of financial statements and tax returns to determine whether a taxpayer has correctly reported their tax liabilities. 
In order to tackle tax evasion and avoidance effectively by conducting tax audits it is essential to identify taxpayers with a substantial probability to evade tax or to assist any other person or company involved in tax evasion. In this regard, the Austrian Ministry of Finance \emph{inter alia} applies predictive analytics as a means to identify the ``right'' target company for tax audits. The ``right'' target company in this context means each company less inclined to comply with tax rules and thus with a high risk or probability, respectively, for tax avoidance. 
We refer to Holy et al.~\cite{holy20} for further information on this matter.

Due to the fact that the task-technology fit describes the interdependence between a technology user, technology itself and the task which needs to be performed, it is essential to define the characteristics of the task to be examined. Before doing so it is of great importance to define the persons responsible for this task in order to draw a clear line between the task scrutinized in this paper and the other tasks with regard to tax audits. Hence, this leads to tax auditors, because those are the persons responsible for reviewing and evaluating the accuracy and compliance of tax returns filed by individuals, businesses, and organizations. Their primary task is to ensure that taxpayers are paying the correct amount of taxes based on the relevant tax laws and regulations. Based on the jurisdiction and the complexity of the tax system the specific tasks of tax auditors can vary in different states. However, beside other tasks (e.g. compliance verification, interviews with tax payers, investigations of financial documents, preparation of audit reports etc) the evaluation of risk of tax evasion associated with a particular taxpayer or industry sector can be described as one of the most important tasks, as the tax audit depends on it. After evaluating the risk of tax avoidance and determining a reasonable probability of tax claims in the amount of a minimum of 10 000 EUR \cite{Setnicka:2020}, the taxpayer in question is audited.

This task of risk assessment combined together with tax audits constitutes as well the legally prescribed task of the Austrian tax authority explained above. It is therefore appropriate diving deeper into the task of risk assessment in order to define the concrete task scrutinized for purposes of the task technology fit applied in this paper. In this respect, it should be recalled that for purposes of the task-technology fit a task is defined as the activity carried out by individuals to produce the required output with a given input \citep{goodhue1995task}.  
In view of the above, it has to be dealt with the question, what are the activities carried out by tax auditors in order to decide upon taxpayers worth auditing on basis of the information pool owned by the tax authority.
Specifically speaking, three task components have to be considered: activities, input, and output. The activities carried out consist of financial data analysis, comparisons, interpretation and calculation of a substantial risk (probability) to avoid taxes. The input can be described as the data from financial statements, past tax audits, market analysis, personal data and in general all data collected by the Austrian tax authority. The output is a range of taxpayers with certain tax anomalies leading to subsequent tax claims of 10~000 euros, and thus worth of tax audits. Hence, the selection process of taxpayers to be audited for tax purposes is the task of which the characteristics are described in the following. Since the Austrian tax authority conducts audits in two ares, namely general income tax audits, i.e., audit of income taxes and other relevant taxes, and special value added tax audits~\cite{Gaedke:2013}, i.e., detection of missing trader fraud~\cite{Svetlozarova:2023}, the task characteristics will be described for both types of audits.

\subsubsection{Selection of taxpayers for general income tax audits}
Several considerations play a role in the selection of taxpayers to be audited for tax purposes. On the one hand, it is impossible to audit all taxpayers and tax periods. The tax authority do not have the necessary personnel to do so and a complete audit would contradict the principle of administrative efficiency meaning that the tasks performed by authorities should be appropriate, processed very speedily, incorporate simplicity, and aiming at cost savings. 
On the other hand, there is certainly a significant number of taxpayers that fulfill their tax obligations in accordance with tax law, either of their own guided by moral principles, or because they have been pointed in the right path of virtue by a previous tax audit. However, there is a considerable number of taxpayers not calculating and paying taxes in accordance with tax provisions. In pursuit of its legal mandate to identify those taxpayers and considering the principle of administrative efficiency the Austrian tax authority performs the following sub-tasks (see~\cite{danielmeyer2021aufgaben}).
\begin{itemize}
\item \textbf{Financial data analysis}. Tax auditors analyze financial records, including income statements, balance sheets, ledgers and supporting documents, to verify the accuracy of reported income, expenses, deductions and credits.

\item \textbf{Comparisons with previous data}. After analysing financial data, this data is compared with data from past financial years and data of taxpayers with similar characteristics and within the same industry. 

\item \textbf{Interpretation and identification of discrepancies} This sub-task actually overlaps with the previous step of data comparison. At this stage tax auditors identify discrepancies, irregularities and potential tax avoidance or evasion strategies by comparing the taxpayer's financial records with industry norms, historical data and similar entities. 

\item \textbf{Risk assessment/calculation of probability to avoid taxes}. Tax auditors evaluate the level of risk associated with a particular taxpayer and calculate the probability to avoid taxes. The auditors consider factors such as past compliance history, transaction complexity, tax planning strategies and industry-specific risks. The criterion risk means here that a tax audit will lead to tax liabilities at least in the amount of EUR 10 000. Hence, all taxpayers where such a tax liability was estimated are selected for tax audits. Beside tax auditors, the PACC is selecting taxpayers for tax audits on basis of the risk criterion and by means of predictive analytics-methods, which is further described in Section~\ref{sec:techAspects}. The primary purpose of this procedure is to prioritize audit resources effectively, meaning to select taxpayers with considerable tax liabilities for audits. This was the main reason for introduction of predictive analytics-methods as well. 
\end{itemize}

For the sake of completeness the following selection criteria applied for identification of taxpayers to be audited should also be mentioned, because those affect the above selection process of tax auditors (see~\cite{danielmeyer2021aufgaben} and~\cite[p.~58]{Eberl:2018}).
\begin{itemize}
\item\textbf{Selection on basis of time criterion.} The companies that have not been audited for the longest time are selected for tax audits.
The selection of audit cases is based on the criterion "last year audited". The tax authority's IT system records the year in which a taxpayer was last audited; the decisive factor is not the year in which the audit takes place, but the last year of the audited period. 

\item\textbf{Group selection.} For the selection a mathematical random system (random number generator) is applied. After the time selection has already ensured a selection of audit cases, a computer program developed by the Federal Computing Center selects further companies according to criteria unknown to the taxpayers and tax offices. 

\item \textbf{Individual selection.} The third criterion is based on individual characteristics. In contrast to the other selection criteria, which can be assigned more to general prevention, individual selection primarily pursues the purpose of preventing revenue reductions (special prevention). This may include companies selected for tax audits mandated by the Ministry of Finance, companies not selected on basis of time criterion or group selection, companies selected by the head of the tax administration and/or the head of the tax audit department or even by other tax administrations. The individual selection is carried out on certain criteria which can be called warning signals. These may include massive inconsistencies in tax returns or annual financial statements, serious changes in legal relationships (reorganizations, mergers and acquisitions) or even complaints made by other persons.
\end{itemize}

\subsubsection{Selection of taxpayers for value added tax audits -- missing trader fraud}
Value added tax (VAT) constitutes one essential pillar of government revenues. Due to the possibility of input tax deduction, criminal elements appear on the scene who earn their money with organized input tax fraud. According to estimates by the European police authority (EUROPOL), the most common form of VAT fraud is Missing Trader Intra-Community (MTIC) fraud~\cite{Europol01}. Due to the possibility of input tax deduction the risk of organized input tax fraud is very high and the fight against it remains very challenging for governments. This could be remedied by a system that among other things eliminates VAT offsetting in the entrepreneurial chain. But such an implementation has failed in the past and still remains a vision of future. Therefore, the current VAT system requires extensive control measures. The entrepreneur providing the service/supply must be controlled regarding its payment of VAT, and the recipient of the service/supply must be controlled with respect to its right to deduct input tax.
Control measure constitute three types of VAT controls including a general VAT-audit examining the documentation and notifications of companies regarding their monthly and annual VAT returns in order to determine whether the VAT was calculated in accordance with tax law. Another type of VAT-control is a special VAT-audit based on mathematical and statistical methods which consider risk factors and estimate the risk for VAT-fraud of companies. This risk analysis is carried out on basis of VAT data collected by the tax administration and is transparent, thus a tax auditor receives insights about the risk factors and the reasons why a company was chosen for a VAT audit. This means that in such a case a tax auditor is fully aware of the critical transactions and circumstances in connection with VAT. Last but not least there is another special VAT-audit, monitoring companies neither having an establishment in Austria nor generating revenues connected to real estate property located in Austria. As in mentioned above also in this case the same audit measures are applied: Audit of VAT-returns, risk assessments, examination of notifications, information exchange etc. We refer to Griffioen et al.~\cite{Griffioen01} and Bertl et al.~\cite{Bertl:2019} for more detailed information. 

With regard to the selection of taxpayers for MTIC fraud the same selection process as described above applies, hence taxpayers are selected on basis of the risk criterion. Specifications regarding the risk criterion that are worthy of mention are the following~\cite{OECD2004}.

\begin{itemize}
\item \textbf{Selection of new entries}. Taxpayers that have not declared their VAT-return for more than two years are considered as risky and thus selected for VAT-audits and MTIC fraud in principle.

\item \textbf{General selection on basis of the risk criterion}. Taxpayers that do not fall under the category of new entries are selected provided the financial data analysis leads to substantial tax liabilities.

\item \textbf{Random selection}. Taxpayers are selected by random number generator for the purpose of control of the selection process on basis of the risk criterion described above. Hence, this is a measure to audit the selection process of tax payers itself.
\end{itemize}

\subsection{Technology}
\label{sec:techAspects}
\begin{figure}[ht]
  \centering
  \includegraphics[width=\linewidth]{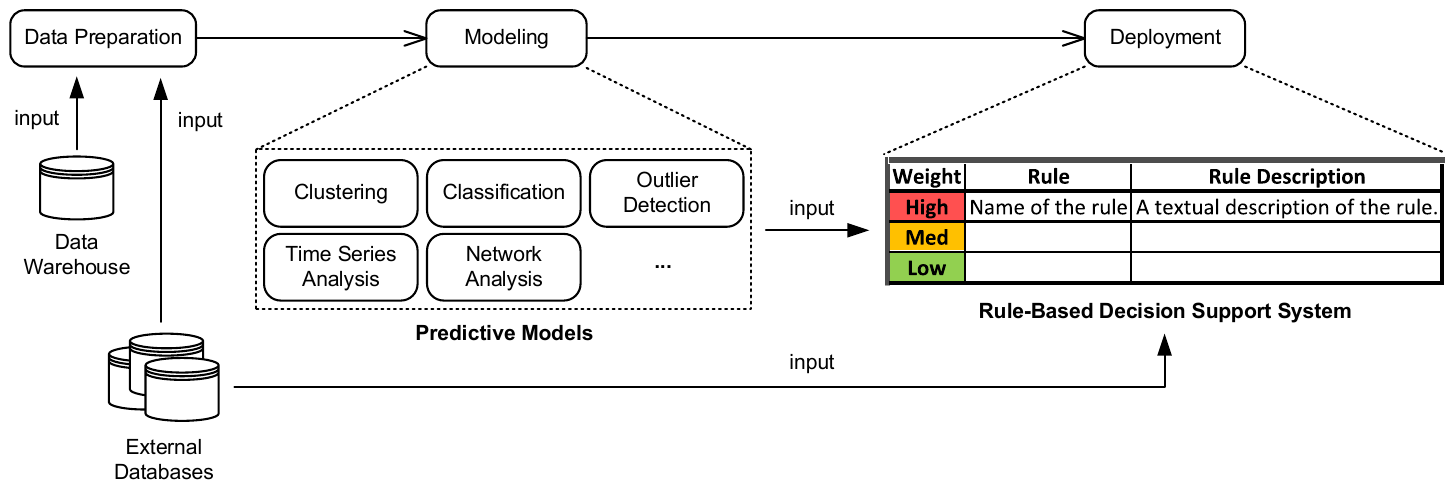}
  \caption{Generic data analysis pipeline of the PACC}
  \label{figure_workflow}
\end{figure}

The Predictive Analytics Competence Center (PACC) follows the Cross-Industry Standard Process for Data Mining (CRISP-DM) for the development of predictive analytics solutions aimed at identifying potential cases of tax fraud.
The CRISP-DM is the de facto standard for conducting predictive analytics projects and consists of six stages -- business understanding, data understanding, data preparation, (predictive) modeling, evaluation, and deployment -- which are conducted in an iterative process (see~\cite{wirth_crisp-dm_2000}).
At PACC, a rule-based decision support system provides the final prediction results, e.g., a ranking of how likely a case might include a fraud.
The rule-based decision support system consists of a set of rules where each rule triggers when a certain condition is fulfilled.
Different rule conditions state different possible fraudulent aspects of a new prediction case.
The input for the rule conditions can include results of predictive models or results of queries over internal and external data sources.
This section outlines the technological aspects of predictive analytics projects performed at PACC and describes a generic analysis pipeline used for predictive case selection demonstrated by the two use cases company audit and missing trader fraud.
Figure~\ref{figure_workflow} shows a graphical representation of the generic data analysis pipeline which is further explained within this section.

Business understanding is the first stage of the CRISP-DM.
General project management, the definition of project goals, and how to achieve those goals are the main activities of this first stage. 
The second and third stage of the CRISP-DM are data understanding and data preparation, respectively. 
At PACC, the data understanding and data preprocessing stages consist of a data scientist extracting and collecting all the required data from the respective sources and load the data into the SAS analysis platform, which serves as the main tool for analysis. 
The collected data include, for example, tax-related data from governmental sources, data from external databases like the Eurofisc (a multilateral warning system of the member states of the European Union for combating VAT fraud), the VIES (VAT Information Exchange System), national databases like the company registry, or transactional data like intra-community acquisitions and deliveries. 
PACC requires a legal foundation to use any kind of data within an analysis. 
If there is no legal foundation for the use of data then its use may be prohibited. 
The legal foundation is further described in Section~\ref{sec:legal_aspects}.
The fourth and fifth stage of the CRISP-DM are modeling and evaluation, respectively. 
The collected and preprocessed data are used as training data to generate predictive models.
At PACC, these models should have the capability to recognize a possible fraud case or a possible conspicuous aspect of a fraud case. 
It is challenging to train only ``one'' big predictive model that is capable to predict all possible tax frauds for all kinds of companies. 
There are two reasons for this.
First, there exists a large number of heterogeneous companies liable to tax.
Second, there are a lot of different fraud approaches to avoid paying taxes. 
Thus, during the modelling stage, PACC data scientists can use a wide range of different prediction models.
Possible modeling techniques include for example clustering, classification, outlier detection, and network analysis.
The result is a hybrid prediction approach including expert rules as well as probabilistic models, where each of the trained models is capable of predicting a part or aspect of a possible fraud case. 
During the evaluation stage, data scientists constantly refine and expand the list of prediction models to be able to react to new fraud patterns.
Given the complexity of the desired prediction problem, ``one'' big model may not be capable to make the final prediction.
Thus, the final prediction may be a combination of the results of many prediction models and further relevant data.
At PACC, a \emph{modular, rule-based decision support system} determines how different prediction models are combined into a final predictions, based on a predefined set of rules, which may include both the results from prediction models based on machine learning as well as assessment based on expert knowledge.

In the rule-based decision support system at PACC, each rule is triggered if the input for the rule fulfills the defined rule condition.
A rule condition may include two different kinds of information. 
First, the rule conditions may depend on the results of prediction models.
For example, a rule can be triggered when a classification model predicts a certain class.
Second, the rule conditions may formalize any kind of expert knowledge.
For example, a rule can be triggered when a specific fraudulent pattern in the company figures occurs which experts know due to years of experience.

The use of a modular rule-based decision support system possesses the following benefits compared to the use of ``one'' big prediction model.
First, the inclusion of new discovered fraud patterns can be easier.
Since training or retraining a single, universal prediction model is not necessary, expert knowledge can easily represent newly emerging fraud patterns in the rule conditions.
Second, the inclusion of rare fraud patterns can be easier.
Rare fraud patterns may not provide enough training data and thus a model can have difficulties to predict these patterns.
Third, the inclusion of many specialized prediction models is possible.
Data scientists can train many smaller models which aim to predict only an aspect of a possible fraud scenario.
Depending on the specific case, not all the rules from the rule-based decision support system may be suitable.
When the rule conditions include inputs that relate to different scenarios than the case under investigation, those rules are not applicable.
For example, a rule condition including the result of a prediction model that was trained on data of companies with completely different characteristics like different branch, company form, number of employees, revenues, etc. 
A data scientist has the possibility to deactivate any rule that is not suitable for each new individual case. 
This avoids triggering unsuitable rules which should not contribute to the final result.

Each triggered rule contributes to the final prediction result to a certain, predefined extent.
This prediction result may, for example, be a \emph{fraudulent score}, ranging from 0 to 999, where 0 indicates a low risk of fraud, 999 indicates a high risk of fraud. 
Each rule has a weight assigned which indicates how much the rule will contribute to the final result. 
This weighting is indicated by a traffic light system (low, medium, high), shown in the user interface, next to the respective rule.
Using this traffic light system within the rule-based decision support system, an user gets an approximate idea which rules contributed how much to the final prediction result.
The assignment of the contributing weights to a new rule is not trivial and can, therefore, be challenging for the data scientists.
Given the heterogeneous structure and diverse input of the rules (some based on expert knowledge, some based on predictive models, etc.), a data scientist assigns the contributing weight individually for each rule. 
For example, a rule can have a small contributing weight if it triggers alone but might have a big contributing weight if it triggers in combination with other rules. 
A data scientist has the option to provide a textual explanation about why a new rule indicates that a case may contain a possible fraud. 
These explanations are then given to the respective auditors if a case is selected for an audit. 
If a rule is based on collected expert knowledge, then this knowledge can be expressed and stored as textual explanation.
If a rule is based on a black box prediction model, then a textual explanation for this rule is often only possible sporadically because it is almost impossible for a human to understand the decision making of the black box model. 

A tax auditor who has an audit case proposed by the PACC will receive additional information, e.g., conspicuous areas in the reports, textual explanations for triggered rules, or predicted potential back payments.
This additional information poses the risk of steering the auditors into a certain direction, thus influencing the auditors' final decisions. 
The PACC, therefore, recommends that auditors, regardless of the additionally received information, conduct a regular audit of the received case, not only addressing the risks raised by the audit case selection. 
The requirement of conducting a regular audit aims at preventing the auditors from focusing only on the raised risk areas while neglecting other areas. 
Nevertheless, any extra information can be helpful because the PACC has a holistic view over not only the given case but also other companies which are similar to the one that is currently audited.
For example, the documents and calculations submitted by a company may appear sound, with the auditor unable to find any problems.
Compared to other companies with similar characteristics, however, the personnel costs might be significantly lower, which signals to the auditor that they should further scrutinize the company in this regard.

The PACC is constantly evaluating and regularly expands the existing prediction rules.
After the selection of a case for a tax audit, the result of the audit is typically available after about half a year, sometimes even longer.
This time-span complicates an immediate evaluation of the prediction approach. 
A metric by which the success of the application of predictive analytics can be measured is the success rate (how many of the predicted fraud cases were then actually frauds) of fraud cases found by the PACC's case selection compared to the success rate of cases selected by other criteria.
Yet, it must be stated that since the audit requires the collaboration of hints from the predictive analytics tools and the expertise of the auditors, it is not possible to clearly separate these success rates and to fully attribute them to only the application of predictive analytics, which again complicates the evaluation.

In the following, we further describe two of the use cases where the PACC applies the aforementioned rule-based decision support system in practice.

\subsubsection{Predictive Case Selection for Company Audits}
\begin{figure}[ht]
  \centering
  \includegraphics[width=\linewidth]{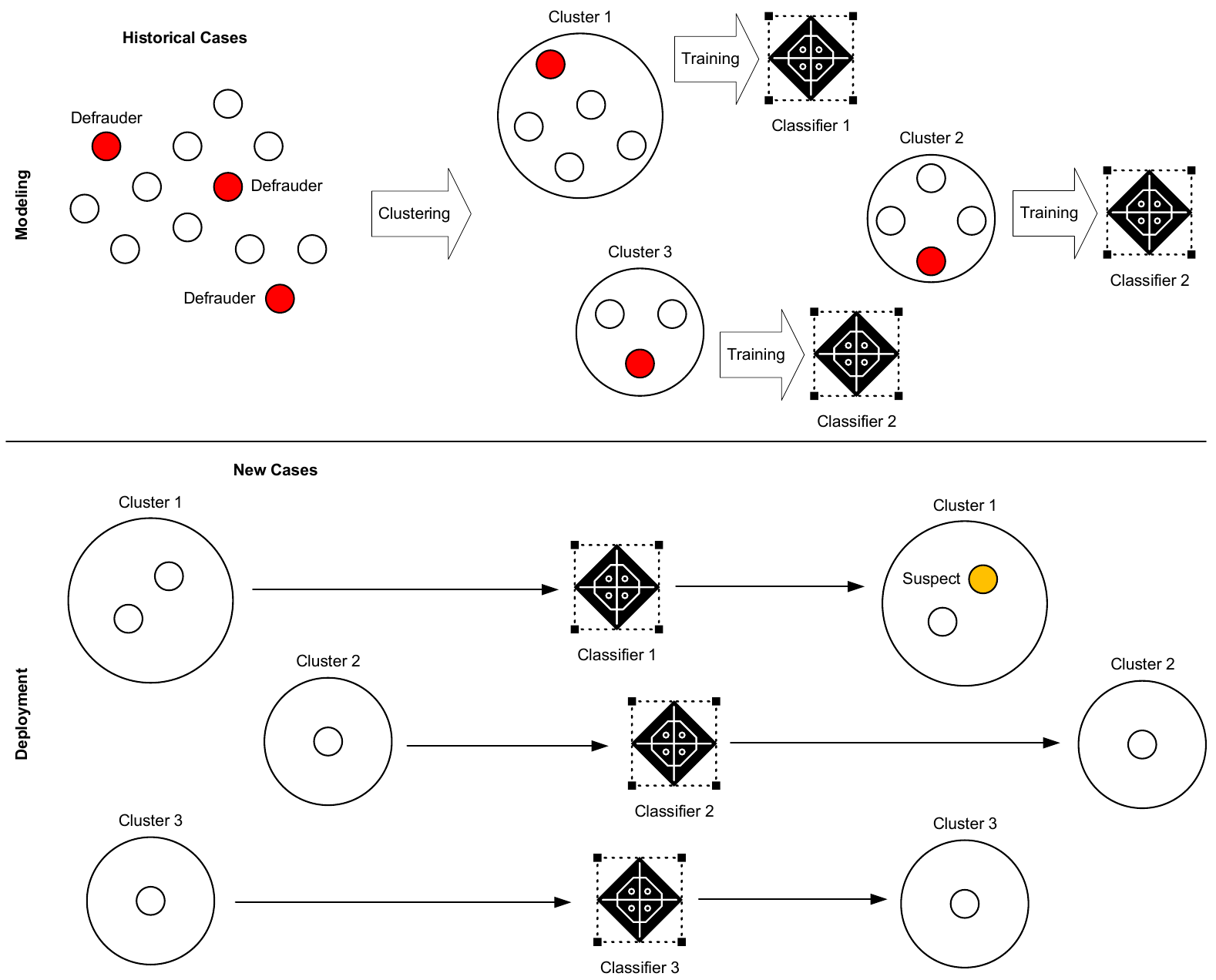}
  \caption{Use of predictive analytics for case selection for company audits}
  \label{figure_companyAudit}
\end{figure}

The predictive case selection for company audits uses a data analysis pipeline which includes a combination of clustering and classification models. 
Figure~\ref{figure_companyAudit} gives an overview of this pipeline. 
The final prediction is solely based on the results of the clustering and classification results. 
Initially, the use of rules based on expert knowledge was also investigated by PACC, but it turned out that these expert rules did not improve performance.
The application area of company audits has already existed for a long time and there is documentation of a large amount of previous audits. 
PACC can use this documentation of previous audits as training data for predictive models. 
Previous audits used for the training concern a vast amount of different companies. 
The first step is to find similar groups of companies within the documented audit cases by use of a clustering model. 
The second step is to train a supervised classification model for each cluster to predict if a new case may contain a fraud. 
New cases are assigned to one of the identified clusters and receive a prediction whether the case could be a potential fraud case from the supervised model that was trained on the training cases from this cluster.
Cases labeled as potential fraud receive a high fraudulence score and will be forwarded to the auditors in order to be examined in detail.

\subsubsection{Predictive Case Selection for Missing Trader Fraud}
The data analysis pipeline which is used to predict missing trader fraud is a hybrid approach composed of different kinds of prediction models using classification, clustering, anomaly detection, social network analysis, international database searches, etc.
The hybrid approach was chosen because it may need to be adapted quickly to represent new fraud patterns which cannot otherwise be identified in a timely manner due to the limited life span of missing trader companies.
If a new fraud pattern is found then it must be easy to integrate into the existing prediction approach, e.g., by adding a new expert knowledge rule representing the pattern instead of first having to collect a sufficient amount of training data to be able to develop a predictive model.
Another reason is that the amount of collected data for some fraud aspects may never be sufficient for the training of a predictive model and, therefore, an approach solely based on machine learning is not feasible to model all given patterns, which also opens the need for a hybrid approach including rules based on expert knowledge and automated database queries.
Table~\ref{figure_rulebase} shows example rules for the missing trader fraud prediction based on different knowledge sources.
The first, third, and fourth rule are simple expert rules, e.g., whether a company employs a person who is linked to another company that is on the Eurofisc watchlist. 
The second rule shown is triggered depending on a risk score which is computed by a predictive model. 
In total, the missing trader fraud detection framework consists of about 150 different rules addressing various aspects, which can give a hint on possible frauds. 
Each triggered rule contributes to the final score assigned to the case according to a certain weight. 
Based on the ranking of the predicted scores for different cases, the cases with the highest scores are forwarded to the auditors for detailed examination. 
Due to the fact that in the context of the missing trader fraud, possible missing trader companies will only exist for a short life span, the cases must be examined promptly.
The scoring of all present cases is accomplished two times per month in order to be able to start an audit in time to react to possible identified anomalies. 

\begin{table}[ht]
\begin{center}
\begin{minipage}{\textwidth}
\begin{tabular*}{\textwidth}{clp{8.6cm}}
\toprule
\textbf{Weight} & \textbf{Rule} & \textbf{Rule Description} \\ \midrule
\cellcolor[HTML]{FE0000}\textsc{High} & Person linked to Eurofisc watchlist & The company has one or more persons linked to another company on the Eurofisc watchlist. \\
\rowcolor[HTML]{EFEFEF} 
\cellcolor[HTML]{FE0000}\textsc{High} & High overall risk & High risk score based on the effectiveness model, qualitative weight and frequency weight. \\
\cellcolor[HTML]{FFCB2F}\textsc{Med} & Few employees & The company has fewer than four employees. \\
\rowcolor[HTML]{EFEFEF} 
\cellcolor[HTML]{34FF34}\textsc{Low} & Multiple address usage & More than 13 companies are located at this address. \\ \bottomrule
\end{tabular*}
\end{minipage}
\end{center}
\caption{Example rules for predicting missing trader fraud}\label{figure_rulebase}
\end{table}

\subsection{Legal Aspects}\label{sec:legal_aspects}

Constitutional states adhere to the \emph{principle of legality}, meaning that government actions have to be grounded in law. 
This principle also applies to Austria's public administration in general and Austrian financial authorities, which are part of Austria's public administration, in particular. 
Authorities' use of automation---including the use of predictive analytics---for tax risk management thus requires a legal basis, which the amendment of Section~114 of the Austrian Law on Tax Procedure (\emph{Bundesabgabenordnung}) provided in 2016. 
The Austrian legislator justified the legislative amendment with a reference to advances in digitization, which necessitate the use of predictive analytics also in tax administration in order to effectively enforce tax equity by preventing and uncovering illegal behavior of taxpayers (see~\cite[p.~12]{explanation_bill_rgv}).
In particular, tax authorities have to employ predictive analytics to efficiently and accurately identify potential cases of tax fraud for audit activities.

The relevant legal provisions allowing Austrian tax authorities to apply predictive analytics are Sections~48d,~114 and~115 of the Austrian Law on Tax Procedure. 
These sections state that tax authorities shall ensure that the state's tax revenues are not unlawfully diminished. 
Moreover, tax authorities shall carefully collect and exchange data required for the assessment of tax cases. 
Austrian law stipulates as well that, for the purposes of data collection, tax authorities shall maintain a registry that stores data about entities subject to taxation, including data about those entities' economic activities.
In particular, Paragraph~4 of Section~114 of the Austrian Law on Tax Procedure entitles Austrian tax authorities to apply automated processing of personal data.
Hence, tax authorities are allowed to inspect a variety of databases by means of automated data processing, including predictive analytics, and to process personal as well as non-personal data obtained from other sources.

While there is a legal basis in Austria for tax authorities to employ means of automated processing of personal and non-personal data for the purposes of risk management and fraud prevention, the \emph{principle of proportionality} limits the extent of such data processing, which must be suitable, necessary, and appropriate for tax authorities to be able to perform their duties.
The objective of the legislation is to allow the use of predictive analytics to effectively and efficiently combat tax fraud by means of predictive analytics. 
In order to achieve this objective, tax authorities are obliged to carefully collect the relevant data for assessment of tax obligations, continuously update and exchange relevant information, prepare and exchange audit reports, investigate taxpayers, and determine the factual and legal circumstances relevant for assessment of tax obligations.
The legal provisions mentioned above can be interpreted in such a way that the tax administration has to pursue its main goal of collecting the taxes required by law in the right amount and at the right time, as well as to analyse and eliminate the risks of not achieving this goal.
All events or circumstances that result in legally owed taxes not being collected or being collected too late are to be identified and prevented.
And this is also the point where predictive analytics plays a key role: The Austrian tax authority believes that nowadays in times of digitisation its main goal and hereby an efficient and targeted risk management can only be realised in a satisfactory way through the implementation of machine learning. So the Austrian provisions for the application of machine learning respectively predictive analytics are worded in a timeless manner and independently of the state of technical possibilities. These provisions also allow the use of the most modern and up-to-date IT and other technical achievements.

\section{Task--Technology Fit}\label{sec:fit}
We employ the task-technology fit (TTF) framework~\cite{goodhue1995task} to examine the fit between technologies employed by the Austrian tax authorities and the tax auditors' tasks in achieving individual performance impacts from provided information technology.
Our study focuses on the main TTF elements: technologies, tasks, and the resulting task-technology fit.
Specifically, we look at the use of information technologies, in our case predictive analytics systems and algorithms applied by PACC, in selecting audit cases with a higher likelihood of tax evasion.
Our analysis covers the systems and algorithms used for missing trader fraud detection and company tax fraud detection.
The individuals in the focus of our study are the state tax auditors, who are responsible for auditing selected cases.
Tax auditors use technologies to pre-select cases, as well as to receive additional information to more accurately identify possible suspicious parts of an audit and thus uncover potential tax evasion. 
In this study, we evaluated the fit of these technologies for the tasks described and the abilities of the individuals, to assess the impact on performance in the field of taxation.
\subsection{Task Characteristics}             
Previous sections described the task of selecting cases with a high risk of tax evasion, especially in the areas of company tax audit and missing trader fraud. 
We summed up the main characteristics describing the selection task by analyzing and investigating how case selection is done by the Ministry of Finance. 
\subsubsection{Purposefulness}
Tax authorities conduct tax audits in order to uncover tax fraud and consequently back taxes, because there is evidence that taxpayers do not act in accordance with tax law and pay taxes in the amount foreseen by tax law. 
The challenge here is to identify taxpayers deserving of the tax audits, meaning individual persons or companies with significant back taxes. So the purposefulness of a tax audit is a substantial criterion, because only tax audits leading to a certain amount of back taxes are considered as purposeful. 
Having in mind that there are millions of taxpayers, it is impossible to audit all taxpayers liable to tax in Austria every year, a pre-selection of audit cases is necessary with the goal to select audit cases with a high risk of fraudulent behaviour. 
This pre-selection is intended to ensure that existing audit resources can be used as efficiently as possible and thus that taxpayers that have anomalies in their financial statements can be audited with priority. 
This procedure is derived from the general principle of economical administration, based on constitutional law, that requires authorities to perform their tasks in line with the standards of efficiency, thriftiness, and convenience. 
All these standards are aimed at optimizing the relationship between the invested resources (input) and the achieved performance (output). 
They are manifestations of one and the same idea, namely a general principle of economy---the \emph{efficiency principle}~\cite{ritz2017verwaltungsokonomie}.

Historically, the pre-selection of audit cases has been conducted mainly by experienced auditors, who tried to identify possible fraud cases by analyzing available taxpayer data before forwarding the most suspicious cases to an audit. 
In this way, the amount of back taxes is to be maximized and thus the maximum possible amount of evaded tax money is to be paid to the state.
Thus, tax authorities shall ensure that tax revenues are not unjustly reduced, or to put it in another way: 
Tax authorities are obliged to fulfill the main legal aim of prevention of tax evasion (section 114 of Austrian Federal Fiscal Code).

Tax evasion can be carried out in more than one way. 
Auditors require experience to recognize different patterns in the data that may indicate possible evaded tax and, therefore, the pre-selection of audit cases is not always trivial. 
Further, depending on the area of the fraud, new fraud patterns are developed and exploited over time. 
Auditors responsible for the pre-selection need to constantly investigate the most recent fraud patterns to be able to recognize them in the provided data. 

\subsubsection{Impartiality}
Tax auditors often rely on their experience and intuition as to which aspects of a taxpayer's case require closer inspection. 
Auditors may be influenced in certain directions, which could affect the aspects to be audited, e.g., aspects being audited for which there is little evidence of tax evasion, as well as the selection of audit cases in the first place, e.g., selecting only taxpayers from a specific industry. 
Based on the Federal Constitutional Act (Bundes-Verfassungsgesetz) Article 7 ``all citizens are equal before the law'', which means that all governmental institutions are required to accord equal treatment to each individual, which also extends to the selection of tax audit cases.
An auditor should select and examine each audit case individually without any unfounded prejudices or external influence.
For example, auditors should not make detailed auditing of personnel expenses dependent solely on the fact that the company operates in a certain industry, e.g., hospitality. 
Auditors should act impartial within their work to treat every individual equal as stated in the law.

\subsubsection{Transparency}
In order for tax authorities to be able to claim additional tax payments after an audit, it is necessary to document exactly which problems in the files handed in by the taxpayer are the basis for the additional tax payments. 
The reason for an additional payment being claimed as well as its amount must be transparent.
An auditor responsible for a case has to prepare the necessary documentation for claiming the additional tax payments.
The authorities cannot claim payments that have not been comprehensibly justified.

\subsubsection{Non-routineness}
Auditors receive a general guideline about how an audit should be performed and which company areas should be addressed.
Due to the huge number of different types of companies, tax audits differ from case to case. 
Thus, an auditor has to consider the individual company and look for possible inconsistencies in the data. 
For example, if a company has significantly lower personnel expenses with the same amount of profit as companies with similar characteristics, this might be a sign of tax fraud. 
Depending on the fraud area, fraud patterns that occur may be constantly changing or new patterns may be exploited through the use of people's creativity. 
An auditor must be flexible and always searching for new possible fraud patterns in order to prevent the widespread of a new pattern.
It does not make sense to spend time auditing an area where the auditor can already determine from an overview that there is no sign of fraud. 
\subsection{Technology Characteristics}
In Section~\ref{sec:techAspects} we describe the technologies that are used for case selection in the area of missing trader fraud and company tax fraud by the PACC.
In the following, we discuss the main characteristics of predictive analytics technologies used for predictive tax audit case selection.
\subsubsection{Accuracy}
There is no point in applying predictive analytics technologies which do not grant a certain level of accuracy.
In this study, we understand accuracy of predictive analytics results as the extent to which the given predictions come true in the real world, that is, whether a predicted tax fraud case can later be identified as an actual tax fraud case which leads to additional tax payments. 
If accuracy of a prediction is not important for the execution of a task then there is often no need for the prediction at all.
In order to assess whether a given application of predictive analytics yields reliable predictions, it is common to assess the performance of these predictions on test data, where the actual result is already known.
To further assess whether this accuracy level, measured on the test data, also holds for new cases, the final results of the audits where fraud was predicted must be known and documented.

\subsubsection{Explainability}
In this study, we understand explainability as the degree to which the process of why a prediction was made, can be explained in a comprehensible human-understandable way. 
Due to the black-box nature of some machine learning methods, for example neural networks, explainability is not always given. 
The use of rule-based prediction methods, such as expert rules or a decision tree (which can be transformed into rules), offer the possibility to explain why a prediction was made.
Depending on whether explainability of a prediction matters or not, different prediction methods may be favorable in terms of explainability but may lack other qualities.
PACC's predictions are made based on rules, which also take into account the predictions of black-box models as shown in Section \ref{sec:techAspects}.
Each of these rules has a description attached which intents to explain why a case was marked for an audit. 
If the basis of such a rule is the result of a black-box model then the explanation for the prediction is often vague.
Current research at the PACC is concerned with how to make black-box predictions explainable, or at least to obtain clues about how to do so and to provide better hints for the auditors.
\subsubsection{Fairness}
Predictive analytics utilizes historic data to predict future events of interest. 
Based on the patterns represented in the historic data, predictive models will identify similar patterns in new cases. 
This poses the risk that possible biases, included in the historic data patterns, are inherited into the predictions and thus these predictions show the same bias again.
In other words, if a model is trained on biased data then chances are high that this bias will also be incorporated into the model and further be reflected in the model's predictions.
Multiple works have already shown the problems which are caused by biased prediction models, in various areas including governments, industry or healthcare~\cite{DBLP:journals/cacm/SrinivasanC21}.
Despite constant efforts to develop models that are as fair as possible, it is very unlikely to find and eliminate all possibilities of bias, thus some bias will probably always remain in the models. 
\subsection{Interactions between Task, Technology, and Individual}
The task-technology fit is characterized by the interactions between task, technology, and the individual performing the tasks. 
Goodhue~\& Thompson~\cite{goodhue1995task} assume that without effort there is no system that provides perfect data and support for complex task requirements. 
In order to assess whether there is a good fit between predictive analytics technology and the task of selecting tax audit cases, we analyze how the characteristics of both, task and technology match using fit factors proposed by Goodhue~\& Thompson. 
We support our findings using statements from our conducted survey and the conducted interviews.
Each statement includes a reference at the end in brackets whether it is taken from the interviews or from the survey. 
The complete survey results are available online~\cite{appendix} and the reference at the end of the statements indicates the column (uppercase letter) and the row (number) where to find the statement in the survey results.

\subsubsection{Quality and Documentation}
Data quality and model quality are some of the most important factors when working with predictions. 
On the one hand it is worth nothing to receive a prediction that does not provide a certain level of accuracy (and in this sense model quality) so that an auditor can rely on the prediction to a certain extend.
Given the probabilistic nature of some prediction methods it is clear that not all predictions will come true, therefore, it is not realistic to be able to achieve 100\% accuracy within all predictions. 
If a system always incorrectly suggests audit cases and, therefore, almost only cases that do not involve fraud are audited, then there is no advantage in using this system over randomly selecting cases. 
This further concludes that the better the accuracy of these predictions, the better the effectiveness of the audits and the more tax refunds can be recovered. 
On the other hand the quality of predictions is highly dependent on the quality of the training data used for the training of the predictive model. 
Auditors are required to document audits that have been performed and may further be used as training data. 
If the quality of this documentation is flawed or incomplete, predictive models trained on this documentation will not be able to correctly predict these fraud patterns in the future. 
From this it can be concluded that the more qualitative the documentation is, the more qualitative the future models will be or how well current models may be improved with new data.
In order to evaluate the accuracy of a model on new data cases, the outcome of these cases must be accurately documented and evaluated to determine if the prediction and also any indication of suspect areas were correct. 
This additional documentation and evaluation work generates additional work for the auditors and increases the time that has to be spent per audit as also indicated multiple times by participants in our survey:
\begin{itemize}
    \item \textit{``there is much more to document'' (A16)}

    \item \textit{``one is forced to document extensively, where one would have seen basically no need without PACC'' (D23)}

    \item \textit{``In addition, there are formal things to consider when concluding with the feedback, etc.'' (C12)}
\end{itemize}

Assessing the accuracy of the predictive analytics technology on new cases requires the actual results of the cases.
As indicated by the following quote from one of our interviews, assessing the accuracy of predictions may only be possible after a significant amount of time has already passed and, therefore, a quick assessment of the accuracy on new cases is often difficult:
\begin{itemize}
    \item \textit{``Feedback on a particular case also takes up to half a year or even longer. This makes improvements and maintenance and optimization much more difficult.'' (interview) }
\end{itemize}

Besides the long delay for the audit results, fraud may also go undetected, and not all aspects of a company can always be checked during an audit. 
If fraud attempts are overlooked, then this can also have an impact on the assessment of the accuracy of the predictions due to a wrong labeling of the case.
\subsubsection{Comprehensibility}
Goodhue's original fit factor \emph{locatability} of data also includes whether the meaning of the data is easy to find out. 
In our study, meaning is important when an auditor receives a case for examination which was suggested by predictive analytics technology. 
When a case is selected for an audit by predictive analytics, the auditors always ask for a reason why this case was marked as suspicious, as also mentioned in our interviews.
\begin{itemize}
    \item \textit{``The first question team leaders always ask is why a company should be audited.'' (interview)}
\end{itemize}
Given the impossibility of auditing all aspects of every taxpayer in Austria, the auditors are interested in understanding the specific data anomalies that prompted the system to predict a case as possibly fraudulent.
In order to satisfy this need, PACC seeks to provide the auditors with additional information and explanations, which are stored together with the rules in the rule base, as indicated in Section~\ref{sec:techAspects}. 
Depending on the triggered rules leading to the selection of the case, the stored information and explanations are handed over to the auditors as starting point for their examination.
The extent to which this information is sufficient to give the auditor enough guidance for the audit depends on the rules triggered and on the rules which can possibly trigger. 
Our survey found that many auditors state to receive insufficient information as to why a case was selected by the system for audit, as for example by the following answers. 
\begin{itemize}
    \item \textit{"the focus of the selection is not very goal-oriented; it is often not possible to understand why the focus of the audit was chosen in the first place; background information on the selection of cases is not communicated" (A17)}

    \item \textit{"the information from the predictive analytics could not always be reconstructed" (B20)}
\end{itemize}

The more closely the provided information aligns with the needs of the auditors, the more efficiently the auditors can perform the audit, resulting in an improvement in their performance.
It should also be noted that the explainability of a prediction can be understood differently from a technical perspective compared to the perspective of the auditors. 
In view of the fact that black-box models often do not provide any clue at all about the selection process, from a technical point of view it is often already a success to get a small hint about the prediction process. 
A small hint, however, may be insufficient for an auditor to obtain a better understanding of the situation for the examination of a case.

\subsubsection{Authorization to Access Data}
The detection of tax fraud necessitates a comprehensive examination of a significant volume of taxpayers' data.
PACC is able to access all government data, both through technical means and through legal authority, as a large portion of fraud patterns are typically discovered within national data.
By utilizing this access, predictive models can be trained on a large amount of training data previously gathered by the government.
Difficulties in obtaining access may occur when fraud patterns transcend national boundaries and involve international companies and data.
Access to foreign data may be constrained by technical feasibility or security concerns.
For example, the missing trader fraud necessitates collaboration among companies from various countries within the EU for it to work.
It was stated in one of our interviews that it is currently not feasible to continuously verify the validity of all company identification numbers of international companies through automation.

\begin{itemize}
    \item \textit{``And, of course, we also check automatically if foreign UIDs are valid. That's something that you have to check with the member states, but there's still room for improvement, because you can only do a limited amount of these checks on a daily basis. Because each member state has put barriers in place so that you don't create a buffer overflow, which you normally have with hacker attacks, and so there's a natural limit there. So you can't query all companies across the EU, unfortunately that's not possible.'' (interview)}
\end{itemize}

Lack of access to necessary data for detecting fraud patterns can negatively impact the performance of predictive analytics. 
To improve this, efforts should be made to ensure full and unrestricted access to all relevant data, both national and international, to enhance the ability of PA to identify all potential patterns and increase the effectiveness of fraud detection.
\subsubsection{Influence/Fairness}
The application of predictive analytics in selecting audit cases can introduce risks in certain situations.
If biased models are utilized for selecting audit cases, it can lead to the selection of cases that reflect these biases, which can compromise the fairness of the selection. 
Auditors must audit all selected cases and thus have no way to escape this bias.
For instance, if a model consistently assigns a high risk score for tax fraud to restaurants solely based on the type of cuisine they serve, without considering other relevant factors, auditors must conduct more frequent audits on these restaurants.
The better a model can detect real fraud patterns, rather than biased patterns, the fairer the selection of audit cases thus found.

Indications from predictions pose the risk that an auditor will only focus on these highlighted areas instead of conducting an independent audit of the case. 
This contrasts with the possibility of discovering additional findings in the data that the auditor would not have looked at in his ordinary audit and thus would have remained undetected, as also exemplified by the following statements from participants in our survey:
\begin{itemize}
    \item \textit{``In some cases, one is guided into test materials that one would not otherwise have seen, in which one is also not so well versed.'' (C7)}

    \item \textit{``one is tempted to work off only the PACC points to complete the case quickly, the initiative could be lost, in my opinion (I think PACC is not bad) there is a danger that especially young auditors rush to evaluations, diagrams, etc. and immediately go into depth without getting an overall impression (overall feeling).'' (C23)}
\end{itemize}

It is important to find a good balance between providing the auditors with enough information to be able to detect possible new fraud patterns and not giving too much information, which may cause the auditors to neglect the remaining part of an audit and only focus on the given areas.
%

\section{Conclusion}\label{sec:conclusion}
The presented study provides valuable insights for both research and practice. 
To answer the question on how predictive analytics is used by Austrian tax authorities for the selection of cases for tax audits, we conducted two interviews with experts of the Predictive Analytics Competence Center (PACC) of the Austrian Federal Ministry of Finance.
Hence, PACC applies a hybrid prediction approach which incorporates probabilistic machine learning models as well as expert knowledge into the final prediction whether a case should be audited.
To answer the question which factors are determining the degree of fit between the selection of cases for tax auditing and predictive analytics technologies, we applied the task--technology fit framework and conducted a qualitative content analysis using data from a survey among tax auditors in Austria.
We analyzed task and technology characteristics and, using the results of the content analysis, we investigated the degree of fit between task and technology.

\subsection{Implications for Research}
Within this study we identified and discussed characteristics of predictive analytics technologies as well as characteristics of the selection of cases for the tax auditing task of government authorities in Austria; we then assessed the fit between task and technology.
Apart from selecting cases for tax audits, other governmental tasks may likewise be supported by the application of predictive analytics technology.
Research may investigate the fit between these respective tasks and predictive analytics technology, using our study as a blueprint and the findings as a reference. 
Depending on the actual task, the task and technology characteristics may differ.
Technology characteristics which contributed positively to the degree of fit in our study may contribute negatively to the degree of fit for other tasks, or do not influence the fit at all.
Future work may investigate similarities and differences between the task and technology characteristics as well as the task--technology fit in our study and that in other applications of predictive analytics technology by government authorities in Austria and other countries.

\subsection{Implications for Practice}
The proposed findings are results of a qualitative analysis of interviews with experts and a survey among tax auditors on the use of state-of-the-art predictive analytics technologies by Austrian tax authorities. 
The described hybrid prediction approach employed by PACC can be used as guideline for other countries who are currently in the early stages of the development of a predictive analytics department aiming to support tax-related use cases.
Furthermore, the findings on the degree of fit between task and technology may provide insights towards increasing the degree of fit of already deployed applications of predictive analytics.

Without high accuracy and reliability, a prediction model cannot be purposefully deployed in practice. 
To achieve the highest possible degree of accuracy and reliability, additional documentation and evaluation work is required, which is first at the expense of the actual task fulfillment, but in the end should generate an advantage through better use of the model. 
The mere prediction, whether an event will occur in the future, may be insufficient in complicated and multifaceted cases, e.g., the selection of cases for tax auditing.

To achieve a performance improvement in the task of audit case selection, a certain degree of comprehensibility of the given prediction for the tax auditor is necessary, even if this comprehensibility cannot be derived directly from the prediction.
Without comprehensibility, tax authorities cannot claim back taxes due to insufficient provable cause. 
If this degree of comprehensibility is not given directly by the model, then an attempt should be made to make this available to the auditors in some other way.
For example, in the rule-based decision support system described in this study, the rules are enriched with additional explanations and information, which ensure a certain degree of comprehensibility of a given prediction for the tax auditor.

The general possibility to collect all input data needed for a prediction does not mean that this is always really feasible. 
Access to necessary input data may be restricted, e.g., in the case of missing trader fraud, where relevant input data are stored in other countries. 
The easier it is to access all necessary input data, the more data can be used as input for the rule-based decision support system, thus increasing the performance of tax auditors.

The issue of bias in predictive models has been extensively discussed in literature.
Governmental institutions should put particular emphasis on the issue of biases in predictions and the (mis)use of possibly biased predictions.
Governmental authorities should take care to ensure that the advantages of utilizing predictive analytics outweigh any potential negative consequences resulting from biased decisions.

\begin{acks}
The authors thank Christian Weinzinger and Matthias Schmidl from the Predictive Analytics Competence Center of the Austrian Federal Ministry of Finance for their time and the interesting insight into their work.
\end{acks}

\bibliographystyle{ACM-Reference-Format}
\bibliography{references}

\end{document}